\documentclass[11pt]{article}
\usepackage{moriond,epsfig}
\usepackage{amssymb}

\def\Journal#1#2#3#4{{#1} {\bf #2}, #3 (#4)}

\def\PLB{{\em Phys. Lett.}  B}

\def\PRD{{\em Phys. Rev.} D}

\def\mco{\multicolumn}

\def\ra{\rightarrow}

\def\ko{K^0}

\def\be{\begin{equation}}
\def\ee{\end{equation}}
\def\bea{\begin{eqnarray}}
\def\eea{\end{eqnarray}}

\begin{document}
\vspace*{4cm}
\title{Neutrino See-Saw Triviality And Lepton Flavour Violation}

\author{Bruce A. Campbell$^1$ and David W. Maybury$^2$\footnote{speaker}}

\address{ $^1$Department of Physics, Carleton University, 
1125 Colonel By Drive, \\ Ottawa ON K1S 5B6, Canada \\ $^2$Rudolf Peierls Centre for Theoretical Physics, University of Oxford, 1 Keble Road,\\
Oxford OX1 3NP, United Kingdom}

\maketitle\abstracts{For the $D=5$ Majorana neutrino mass operator to have a see-saw
ultraviolet completion that is viable up to the Planck scale, the see-saw
scale is bounded above due to triviality limits on the see-saw couplings. 
For supersymmetric see-saw models, with realistic neutrino mass textures, 
we compare constraints on the see-saw scale from triviality bounds, with
those arising from experimental limits on induced charged-lepton flavour 
violation, for both the CMSSM and for models with split supersymmetry.}

\section{Introduction}
The recent observations of neutrino flavour oscillations (see \cite{fy} for review) 
provide compelling evidence for neutrino mass. If we consider the standard model as an effective field theory, the requisite neutrino mass terms can in general be introduced without extension of the standard model field content. From the low-energy point of view, neutrino mass terms appear through a non-renormalizable dimension-five operator constructed from standard model fields, 
\be 
\mathcal{O}_{\nu} = 
{\lambda_{ij}}L^iL^jHH + \mathrm{h.c.} 
\label{nu_op} 
\ee 
where $i$ and $j$ refer to generation labels, and $\lambda_{ij}$ define arbitrary coefficients with dimensions of inverse mass. The experimental lower bound $\Delta(m^{2}) > 0.0015 (\mathrm{eV}^{2})$ for the oscillation $\nu_{\mu} {\not\rightarrow} \nu_{\mu}$ implies that in a mass-diagonal basis for the neutrinos, and with $\langle H \rangle = 250/{\surd{2}}$, there exists a neutrino mass eigenstate for which (diagonal) $\lambda > 1.6 \times 10^{-15} (\mathrm{GeV}^{-1})$.

Because $\mathcal{O}_{\nu}$ is an irrelevant operator, we may use unitarity constraints on the cross-section for $\nu\hspace{1mm} H \rightarrow \nu^c \hspace{1mm} H$ for the largest neutrino mass eigenstate to place an upper bound on the scale at which the neutrino mass operator $\mathcal{O}_{\nu}$ must be replaced by its underlying UV completion. As $\mathcal{O}_{\nu}$ is a local operator, the scattering process occurs in the s-wave, and we have a partial-wave unitarity constraint on the cross-section. Comparing this to the Born approximation for the cross-section we find that:
\be
\sigma = \frac{2}{\pi}\left({\lambda_{ij}}\right)^2 
\leq 
\frac{8\pi}{s}.
\ee
Since the large $\Delta(m^{2})$ from the atmospheric $\nu_{\mu} {\not\rightarrow} \nu_{\mu}$ oscillations established $\lambda > 1.6 \times 10^{-15} (\mathrm{GeV}^{-1})$ for the largest mass eigenstate, partial wave unitarity for scattering of that mass eigenstate requires that:
\be
\label{UV_complete}
\sqrt{s} \leq 3.8 \times 10^{15} \mathrm{GeV}. 
\ee
Therefore the dimension-five neutrino-mass operator, $\mathcal{O}_{\nu}$, requires its incorporation in a UV complete theory at an energy scale no higher than that given by eq.(\ref{UV_complete}).
The relatively large neutrino mass squared differences obtained from the atmospheric oscillation results of Super-Kamiokande, and the K2K experiment, unambiguously indicate a breakdown of the standard model description of neutrino mass physics at a sub-Planckian (but typically high) energy  scale. While eq.(\ref{UV_complete}) provides an upper bound on the scale of the UV completion, it does not specify a lower bound.

The see-saw mechanism provides, perhaps, the most elegant UV completion that generates the operator $\mathcal{O}_{\nu}$. By itself the see-saw mechanism does not necessarily provide a fully viable UV completion all the way to the Planck scale. The see-saw mechanism involves Yukawa interactions which are not asymptotically free. If the mass scale of the see-saw is sufficiently large, the initial size of these Yukawa couplings at the see-saw scale will be large enough that, under renormalization group evolution in the see-saw extended standard model, the Yukawa couplings will be driven to a Landau pole at energies below the Planck scale. 
By requiring that the see-saw mechanism does not suffer a Landau pole below the Planck scale, an upper bound can be established on the scale at which the see-saw is introduced, and that the corresponding bound is stronger than that following from partial-wave unitarity alone \cite{cm}.  

These considerations become of particular interest in supersymmetric extensions of the standard model where the soft supersymmetry-breaking are introduced at the Planck scale.
In the presence of neutrino masses, our considerations above indicate that we need to introduce new fields, such as the see-saw singlets, to retain the possibility of a viable theory up to the Planck scale. But this implies an interval of RGE running of the soft masses between the Planck and see-saw scales. This running can induce flavour violation in the soft scalar mass-squared terms which can induce charged-lepton flavour violation. Experimental limits on LFV will (depending on the size of the low-energy superpartner masses) give us upper bounds on the size of the Yukawa superpotential interactions involved. The Yukawa couplings in turn are responsible for the neutrino masses via the see-saw, so the observed neutrino masses will give us upper bounds on the see-saw scale. It is now interesting to determine whether we arrive at stronger bounds on the scale of the neutrino see-saw from the requirement that we avoid a Landau pole below the Planck scale (the ``triviality" bound), or from the requirement that the induced charged-lepton flavour violation not exceed present experimental limits. 
In particular, we will examine the low-energy impact of supersymmetric see-saw (non)-triviality, on charged-lepton flavor violation in parameter ranges of the CMSSM, and split supersymmetry 
consistent with all laboratory and cosmological observations \cite{cm}.

\section{Results}

We consider both the CMSSM and split supersymmetry \cite{ark} with see-saw Yukawa couplings that saturate their perturbative limit at the Planck scale. This will have the effect of pushing the Landau pole beyond the Planck mass, while ensuring the largest possible see-saw scale. We find that the upper bound on the see-saw scale becomes, $M = 1.2 \times 10^{15}\hspace{1mm}\mathrm{GeV}$, approximately a factor of three below the naive scale of violation of partial-wave unitarity. 
In both studies, we separately consider strongly hierarchical right-handed neutrinos and degenerate right-handed neutrinos, assuming a normal left-handed neutrino hierarchy with LMA solution. In hierarchical right-handed neutrino case, the see-saw Yukawas depend on a high-energy input parameter dentoed as $\theta_1$ \cite{ca}. 

\begin{figure}[ht!]
   \newlength{\picwidthb}
   \setlength{\picwidthb}{3in}
   \begin{center}
       \resizebox{\picwidthb}{!}{\includegraphics{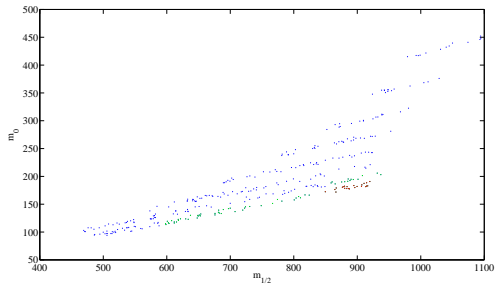}}
   \end{center}
   \caption{Approximate CMSSM region left after LFV constraints imposed, 
$\tan\beta=5$--$40$. Percent of $\theta_1$ parameter space allowed: 
Red $25\%$ -- $20 \%$, green $20\%$ -- $10\%$, 
blue $10 \%$ -- $5 \%$; $\mu >0$}
   \label{percent}
\end{figure}

In the CMSSM with right-handed neutrinos, the soft terms can induce lepton flavor violation through sfermion flavor mixing generated by radiative effects. Presently, the strongest bounds on charged-lepton flavour violation, within the model class that we consider, come from $\mu\rightarrow e\gamma$ \cite{cmm},\cite{jm}. If we consider the allowed CMSSM parameter space, consistent with all laboratory and cosmological constraints \cite{ellis}, we find that most of angle $\theta_1$ becomes eliminated for a see-saw scale of $\Lambda=1.2 \times 10^{15}\hspace{1mm}\mathrm{GeV}$. Figure \ref{percent} demonstrates the constraints on $\theta_1$.
If we assume a degenerate spectrum of right-handed neutrinos the dependence of $\mathrm{BR}(\mu\rightarrow e\gamma)$ on the angle $\theta_1$ is lost. In this case, the prediction for the branching ratio is fixed by the low energy parameters of the PMNS matrix, the light neutrino masses, and the see-saw scale. We find that the entire range of CMSSM parameter space is eliminated in the degenerate $\nu_R$ case.

In models of split supersymmetry \cite{ark}, the gaugino and higgsino mass spectrum remains light (TeV range) and the A-terms remain small, protected from large masses by $R$-symmetry, while the sfermions acquire large (from multiple TeV to GUT scale) masses at the supersymmetry breaking scale. As the CMSSM case demonstrates, the branching ratio rises sharply for large see-saw Yukawa couplings (or equivalently large see-saw scales). If large see-saw Yukawa couplings are assumed, scalar mass terms require large values in order to drive the branching ratio for $\mu\rightarrow e\gamma$ down to acceptable levels. We determine the scale of scalar masses that can be probed with LFV in this set-up. We assume a split supersymmetry scenario with large universal scalar masses and light universal gaugino and higgsino masses imposed at the Planck scale.

\begin{figure}[ht!]
\newlength{\picwidthc}
\setlength{\picwidthc}{2.2in}
 \begin{center}
{\resizebox{\picwidthc}{!}{\includegraphics{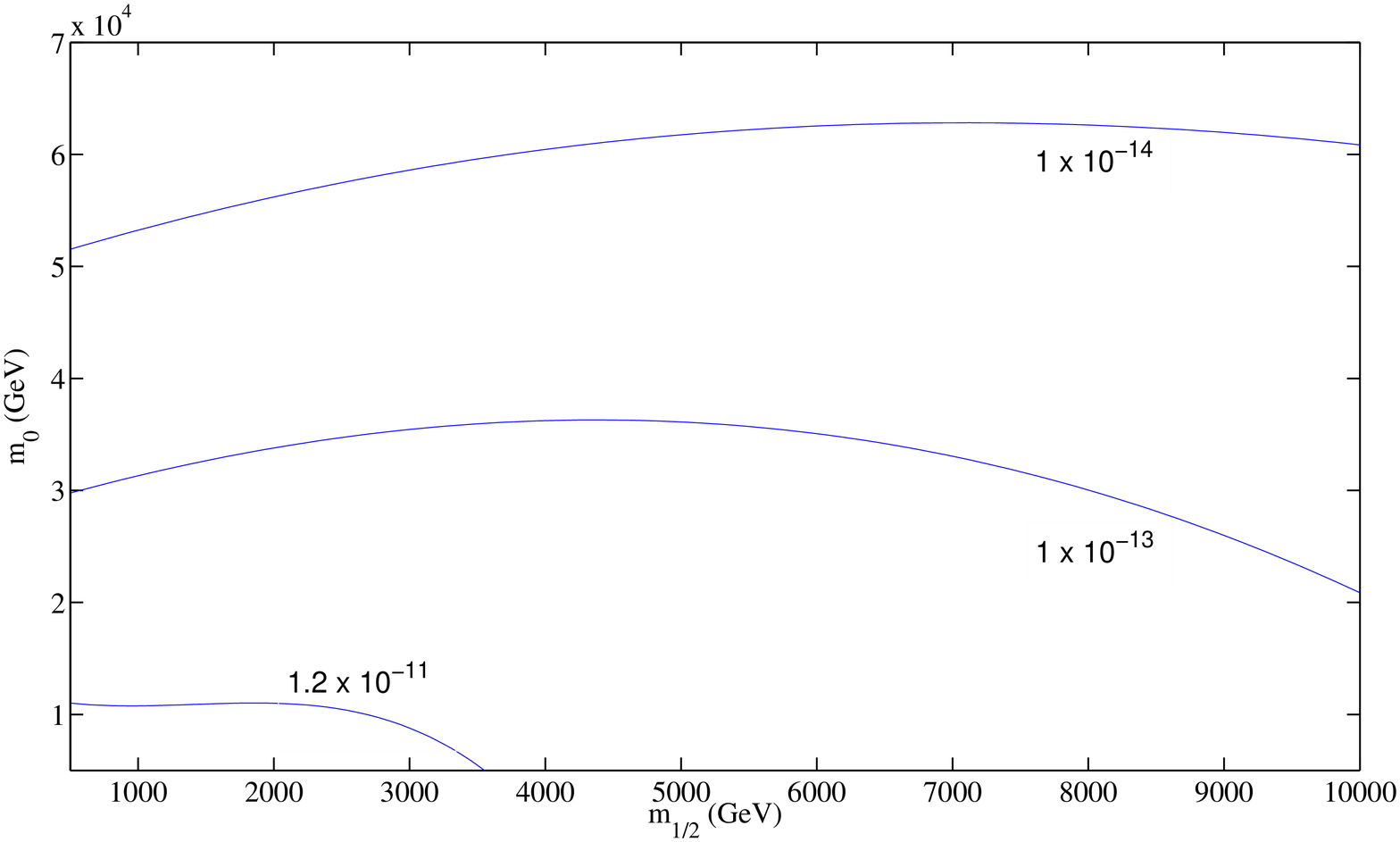}}}
{\resizebox{\picwidthc}{!}{\includegraphics{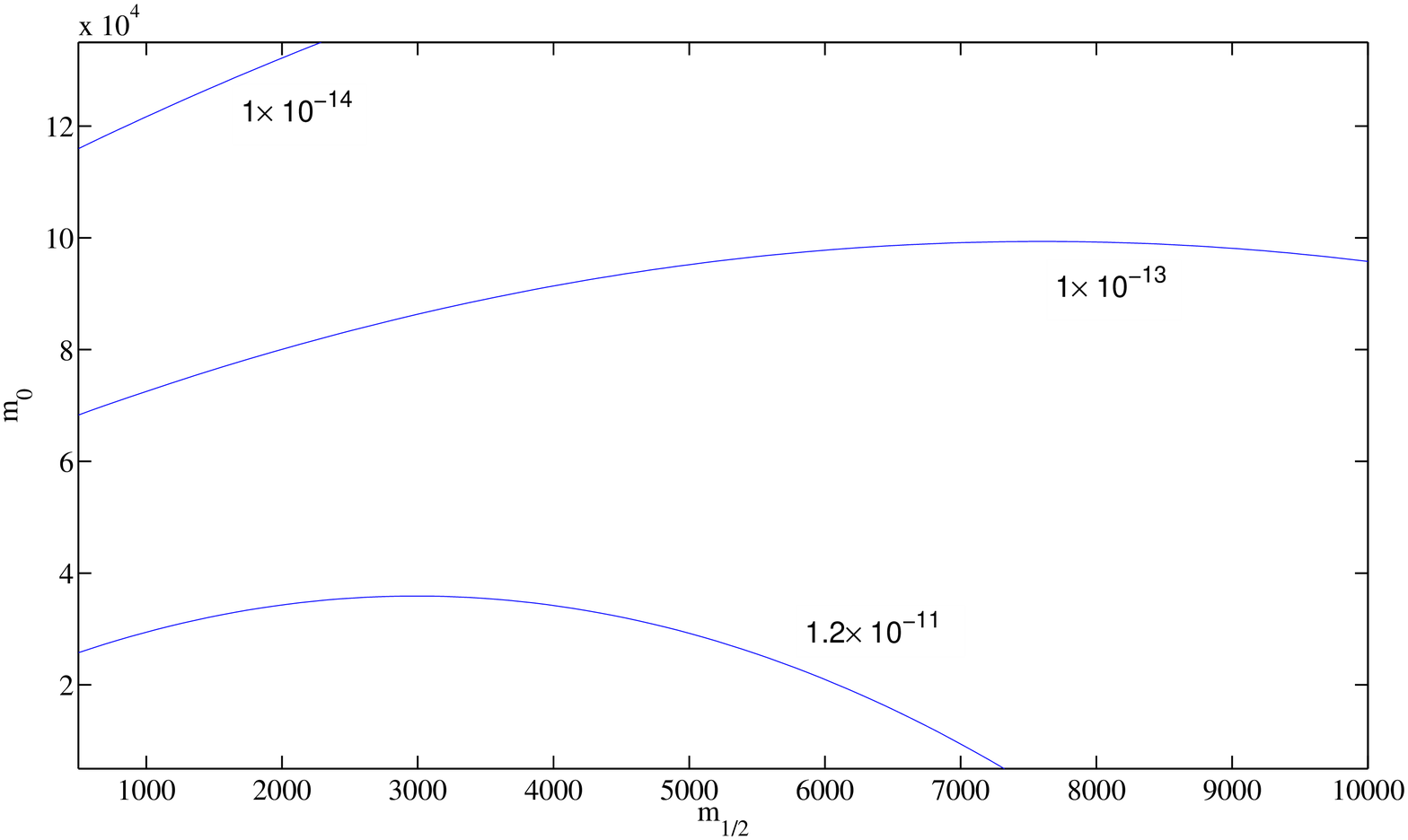}}}
 \end{center}
\caption{Contours of constant branching ratio, $m_{1/2} =1000$ GeV, 
$\mu >0$, $\nu_R$ hierarchical;  $\tan\beta=5,40$ with $\theta_1=1.4$}
\label{tanb1}
\end{figure}

\begin{figure}[ht!]
\newlength{\picwidthh}
\setlength{\picwidthh}{2.2in}
 \begin{center}
{\resizebox{\picwidthh}{!}{\includegraphics{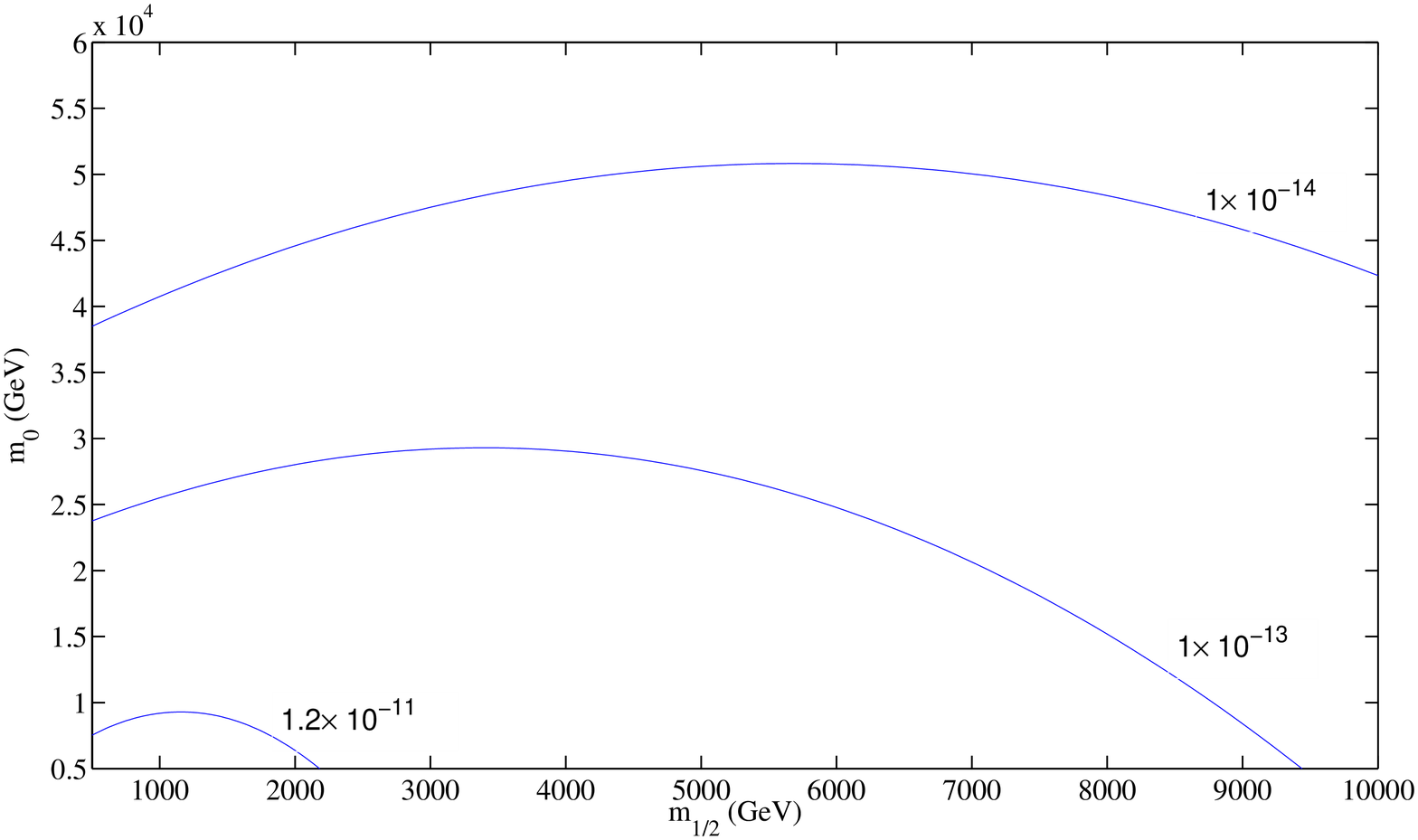}}}
{\resizebox{\picwidthh}{!}{\includegraphics{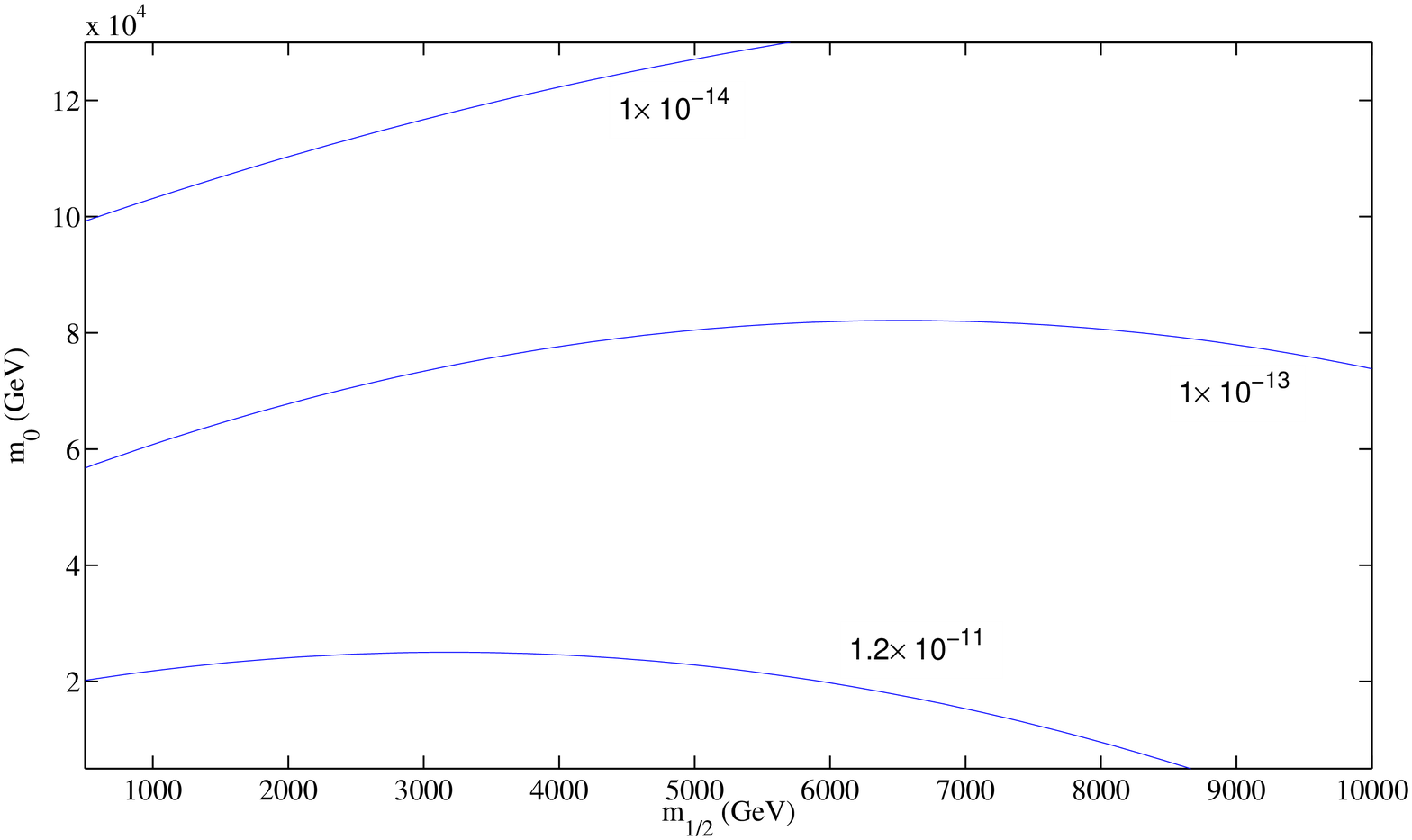}}}
 \end{center}
\caption{Contours of constant branching ratio, $m_{1/2} =1000$ GeV, 
$\mu >0$, $\nu_R$ degenerate;  $\tan\beta=5,40$.}
\label{tanb2}
\end{figure}

From figures \ref{tanb1} -- \ref{tanb2} we find that LFV can probe scales as high as $\sim 100$ TeV in this scenario. In the cases with hierarchical right-handed neutrinos we choose $\theta_1 = 1.40$, exploring a region of parameter space that corresponds to the maximum amount of LFV. It should be noted that $0.8<\theta_1<1.5$ corresponds to LFV levels within one order of magnitude the maximum value. 
Note that the graphs indicate a branching ratio at the $1\times 10^{-13}$ level with universal scalar mass approaches $50$ TeV in most cases. 
Upcoming searches of the MEG experiment 
expect to explore BR($\mu\rightarrow e\gamma$) at the $1 \times 10^{-13}$ level which, as indicated from figures \ref{tanb1}--\ref{tanb2}, corresponds to scalar masses $30$ -- $80$ TeV range. The $100$ TeV range is of particular interest as bottom up considerations have shown \cite{wells} that PeV scale ($10^3$ TeV) split supersymmetry presents a highly suggestive phenomenology   
In these models, much of the parameter space yields scalar masses in the $100$ TeV range. 

\section{Conclusions}

By requiring the absence of a Landau pole in the see-saw sector below the Planck scale (``triviality" bounds), we determined bounds on the see-saw scales in supersymmetric extensions of the standard model. Conservatively, we demanded that the see-saw Yukawa couplings remain perturbative up to the Planck scale, establishing an upper bound on the see-saw scale of $1.2\times10^{15}$ GeV in the supersymmetric case. See-saw scales near these bounds lead to large Yukawa couplings over the domain of renormalization group running up to the Planck scale which in the supersymmetric case, with gravity-mediated supersymmetry breaking, leads to strong constraints from lepton flavour violation.

In the CMSSM, most of the allowed parameter range is eliminated for a see-saw scale of $1.2\times10^{15}$ GeV with hierarchical $\nu_R$s and completely eliminated in the degenerate $\nu_R$ case. If the LHC establishes the CMSSM, the low level of charged-lepton flavor violation implies a see-saw scale much below the bounds given by triviality considerations for most 
of the see-saw parameter range. However, in supersymmetric models with large scalar masses, such as split supersymmetry, lepton flavor violation can be used as a probe of see-saw scales near the triviality limit. If $\mu\rightarrow e\gamma$ is observed at levels that can be probed in the MEG experiment, and if a moderately split supersymmetric spectrum is realized with see-saw generated neutrino masses, the implications would include a high see-saw scale near its triviality bound.

\end{document}